\documentclass{article}

\usepackage{graphicx}
\usepackage{psfig}
\usepackage{epsfig}
\usepackage[round]{natbib}

\title{Improving probabilistic weather forecasts using seasonally varying calibration parameters}

\author{Stephen Jewson\footnote{\emph{Correspondence address}: RMS, 10 Eastcheap,
London, EC3M 1AJ, UK. Email: \texttt{x@stephenjewson.com}}\\
RMS, London, United Kingdom}
\begin{document}

\newcommand{\bx}[1]{\fbox{\begin{minipage}{15.8cm}#1\end{minipage}}}

\maketitle

\maketitle
\begin{abstract}
We show that probabilistic weather forecasts of site specific temperatures can
be dramatically improved by using seasonally varying rather than constant calibration parameters.
\end{abstract}

\section{Introduction}

Different users of weather forecast are interested in different things. 
One particular group of users, including weather derivatives traders, 
is most interested in probabilistic forecasts at specific locations. 
The production of such forecasts on the 0-10 day timescale
is what we will consider in this article.
We will derive our forecasts from the output
of numerical weather prediction models, as is usual. 
This output contains a lot of information about the future weather
but needs processing in order to be directly relevant to individual sites, which are
not represented in the models. 
This processing step, usually known as \emph{calibration} or \emph{downscaling} 
can usefully be cast as a 
classical (i{.}e{.} non-Bayesian) statistical problem. 
In this framework a mixture of climatology and the output 
from the model are used as predictors and
the observations that we wish to predict are the predictand.

Forecast calibration can be performed for any weather variable,
but we will focus on temperature.
For temperature anomalies\footnote{from which the mean
and seasonal cycle have been removed}
a good starting point for the calibration model is a standard linear regression taking the ensemble mean
as the single predictor, and the distribution of possible future temperatures as the predictand.
Linear regression has been used as a calibration model for at least 30 years, although it is only recently
been fully appreciated that it gives a good probabilistic forecast, rather than just a good forecast of
the expected temperature.
The challenge is now to build models
that perform better than linear regression, and that is the subject of this article. We have
described the testing of a number of models versus linear regression in previous articles, 
and have generally found it hard to beat by more than minute amounts.
For instance, we have tried to improve forecast skill by using the ensemble spread as a predictor of uncertainty
(\citet{jewsonbz03a} and~\citet{jewson03g}),
but found only a small benefit. We have also tried to improve the forecast by relaxing the assumption of
normality and replacing the normal distribution with a kernel density, 
but in that case found almost no benefit at all~\citep{jewson03h}.

In another study we investigated whether the benefit of using an ensemble versus a single forecast arises
more from the information content of the ensemble mean or the ensemble spread~\citep{jewson03i}. 
The answer is very clear: the ensemble mean
is vastly more useful. This suggests that the best way to beat linear regression might be to improve
the forecast of the mean, rather than the forecast of the spread, and that is the approach we
follow below. 

How could we improve the forecast of the mean? 
There is a long list of methods we might consider, including:
\begin{itemize}
  \item using non-linear models such as neural nets
  \item using predictors from other locations
  \item using lagged predictors
  \item using multiple models
  \item using seasonally varying parameters
\end{itemize}

These are all probably worthy of investigation. 
However in this paper we will only address the last of these:
can we beat constant parameter linear regression by allowing the parameters to vary seasonally?
On the one hand, from a meteorological point of view, this seems a very reasonable approach since
in the climate one usually finds that everything varies seasonally. On the other hand,
as devotees of parsimony, we balk at this approach. 
In a seasonal parameter model each of the parameters in the regression
becomes (at least) 3 parameters.
Thus a 3 parameter model becomes a 9 parameter model. For such a model to be better, the seasonality
in the mapping from model forecast to observed temperature had better be rather strong.

The idea of using seasonally varying parameters is somewhat similar to a method currently 
used by some National Meteorological Services for
the calibration of single forecasts which builds the calibration models using only recent training data 
(e{.}g{.} data for the previous 90 days). This method automatically captures some aspects of seasonality because it allows the
calibration parameters to vary through the year. However, we believe that explicitly modelling seasonality
has some benefits, and will be the method of choice in the long run. This is because:

\begin{itemize}
  \item Fitting calibration parameters from the previous 90 days suffers from the problem that the parameters
  are always slightly behind relative to the present point in the season. They may be the best parameters for
  45 days ago, but will not be the best parameters for today.
  \item Fitting calibration parmaeters from the previous 90 days only allows us to use a small amount of 
  past forecast data for estimating the parameters. As the amount of available past forecast data increases
  it makes sense to try and use all of this data. This is 
  especially important if we are to make use of the subtle signals contained in the varying spread of
  ensemble forecasts.
\end{itemize}
 
In section~\ref{data} we discuss the data we use for our study. In section~\ref{models} we describe the models
and how we will compare them. In section~\ref{results} we present the results, and in section~\ref{conc} we present 
our conclusions and discuss areas for future work.

\section{Data}
\label{data}

We will base our analyses on one year of ensemble forecast data for the weather
station at London's Heathrow airport, WMO number 03772. The forecasts are predictions
of the daily average temperature, and the target days of the forecasts
run from 1st January 2002 to 31st December 2002. The forecast was produced
from the ECMWF model~\citep{molteniet96} and downscaled to the airport location using a simple
interpolation routine prior to our analysis. There are 51 members in the ensemble.
We will compare these forecasts to the quality controlled climate
values of daily average temperature for the same location as reported by the UKMO.

There is no guarantee that the forecast system was held constant throughout this period,
and as a result there is no guarantee that the forecasts are in any sense stationary,
quite apart from issues of seasonality. This is clearly far from ideal with respect to 
our attempts to build statistical interpretation models on past forecast data but is,
however, unavoidable: this is the data we have to work with.

Throughout this paper all equations and all values are in terms of double anomalies
(have had both the seasonal mean and
the seasonal standard deviation removed). 
Removing the seasonal standard deviation
removes most of the seasonality in the forecast error statistics, and partly justifies the use of
non-seasonal parameters in the statistical models for temperature that we propose.

\section{Models}
\label{models}

As mentioned in the introduction, we will take a \emph{classical statistical} approach to the problem of creating
probabilistic temperature forecasts. 
This means that we will postulate models which predict the distribution
of temperature directly in terms of a number of predictors. Such models are simple to design, 
simple to understand, simple to fit, simple to test and easy to use for making forecasts. 
They also allow us to incorporate any number of predictors, 
including climatology, in an optimum (likelihood maximising) way.

The standard method for estimating the parameters of classical statistical models is to
find those parameters which maximise the probability of the observations given the model. 
This quantity, when considered as a function of the model parameters, is known
as the likelihood. A standard (and intuitively very reasonable) way to compare such models against
each other is to compare the maximum likelihoods they achieve. 
The forecast that gives the higher likelihood (or log-likelihood)
is the better forecast (see~\citet{jewson03d} and~\citet{jewson03f} for more details on this). 
This method for comparing probabilistic forecasts can
be used on both continuous and discrete forecasts, and in-sample or out-of-sample. In-sample testing has two
caveats: firstly that it can only be applied to parsimonious parametric models 
(because non-parametric models and models with large numbers of parameters are overfitted)
and secondly that the probability used in the
comparison has to be adjusted to penalise models with more parameters. A common way to make this adjustment is to
use the AIC criterion, and the AIC score is what we will use below.

We now present the models we compare in this study.

Our first model is simply linear regression between temperature on day $i$ ($T_i$) and the ensemble
mean on day $i$ ($m_i$), which we write as:

\begin{equation}\label{regression}
  T_i \sim N(\alpha_0+\beta_0 m_i, \gamma_0)
\end{equation}

This model corrects biases using $\alpha_0$, optimally "damps" the variability of the ensemble mean 
and merges optimally
with climatology using $\beta_0$, and predicts flow-independent uncertainty using $\gamma_0$.
The bias and the uncertainty produced by this model vary seasonally because of the deseasonalisation
and reseasonalisation steps.
All our subsequent models will be judged against this model.

Our second model generalises this model so that the parameters themselves vary seasonally:

\begin{equation}\label{sregression}
  T_i \sim N(\alpha_i+\beta_i m_i, \gamma_i)
\end{equation}
where
\begin{eqnarray}
  \alpha_i&=&\alpha_0+\alpha_s \mbox{sin} \theta_i+\alpha_c \mbox{cos} \theta_i \\\nonumber
  \beta_i &=&\beta _0+\beta_s  \mbox{sin} \theta_i+\beta_c  \mbox{cos} \theta_i \\\nonumber
  \gamma_i&=&\gamma_0+\gamma_s \mbox{sin} \theta_i+\gamma_c \mbox{cos} \theta_i
\end{eqnarray}

where $\theta_i$ is the time of year. 
We have represented seasonality in the simplest way possible by using just one harmonic in
order to keep the number of parameters as low as possible.

We now consider three models that are intermediate between the constant-parameter linear 
regression (equation~\ref{regression}) and the seasonal-parameter linear regression (equation~\ref{sregression}).

The first only has seasonal bias correction:
\begin{equation}
  T_i \sim N(\alpha_i+\beta_0 m_i, \gamma_0)
\end{equation}

The second has seasonal damping:
\begin{equation}
  T_i \sim N(\alpha_0+\beta_i m_i, \gamma_0)
\end{equation}

and the third has seasonal innovations:
\begin{equation}
  T_i \sim N(\alpha_0+\beta_0 m_i, \gamma_i)
\end{equation}

For comparison we will also consider the spread regression model of~\citet{jewsonbz03a}:
\begin{equation}
  T_i \sim N(\alpha_0+\beta_0 m_i, \gamma_0+\delta_0 s_i)
\end{equation}

(where $s_i$ is the ensemble spread on day $i$)
and a completely seasonal version of the spread regression model:

\begin{equation}
  T_i \sim N(\alpha_i+\beta_i m_i, \gamma_i+\delta_i s_i)
\end{equation}

where

\begin{eqnarray}
  \alpha_i&=&\alpha_0+\alpha_s \mbox{sin} \theta_i+\alpha_c \mbox{cos} \theta_i \\\nonumber
  \beta_i &=&\beta _0+\beta_s  \mbox{sin} \theta_i+\beta_c  \mbox{cos} \theta_i \\\nonumber
  \gamma_i&=&\gamma_0+\gamma_s \mbox{sin} \theta_i+\gamma_c \mbox{cos} \theta_i \\\nonumber
  \delta_i&=&\delta_0+\delta_s \mbox{sin} \theta_i+\delta_c \mbox{cos} \theta_i
\end{eqnarray}

This last model has the greatest number of parameters of all the models we consider (12) and would
be expected to perform the best because it includes all the other models.

We fit the parameters of all the models by maximising the likelihood using a standard quasi-Newton method
with finite-difference gradient.

\section{Results}
\label{results}

\begin{figure}[!htb]
  \begin{center}
    \includegraphics{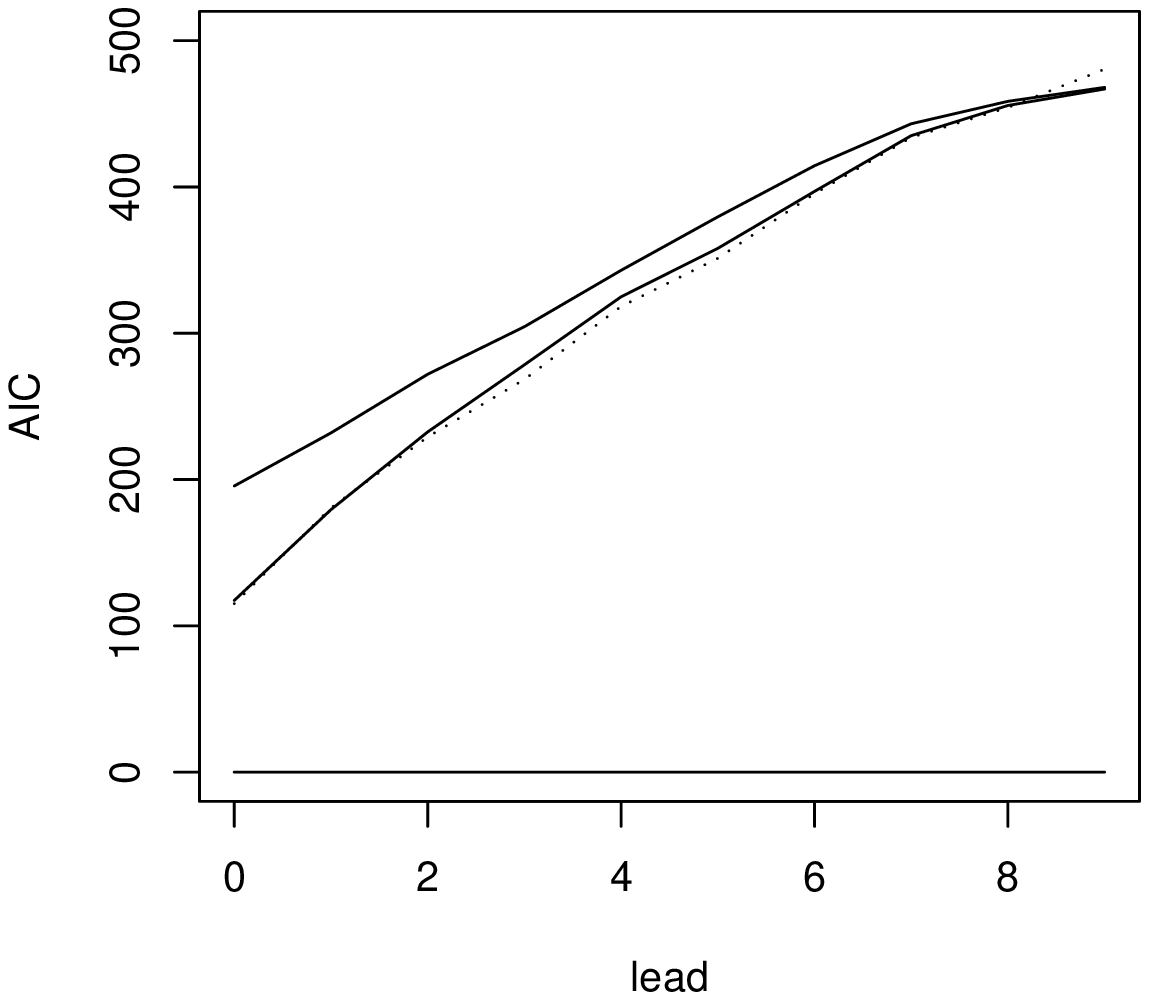}
  \end{center}
  \caption{
The AIC scores for probabilistic forecasts made using linear regression (top solid line) and
linear regression with seasonal parameters (lower solid line). The dotted line
shows spread regression with seasonal parameters. Low scores are better.
          }
  \label{f:f1}
\end{figure}

Figure~\ref{f:f1} shows the AIC scores for the constant parameters and seasonal parameter linear
regression models (upper and lower solid lines respectively). By definition, lower values of the AIC
score are better, and a value of zero would be a perfect forecast. We see a \emph{vast} improvement
in the skill of the probabilistic forecast as a result of using seasonal parameters, especially at
short lead times. The dotted line shows the AIC score for the seasonal spread regression model.
There is a further improvement at most lags from making the spread calibration parameters seasonal in addition, 
but it is small. At lag 9 the fitting of the parameters of the spread regression failed: the algorithm
was unable to find a convincing maximum for the likelihood. This is presumably because the signal is too
weak given the number of parameters we are trying to fit and the amount of data being used. 

\begin{figure}[!htb]
  \begin{center}
    \includegraphics{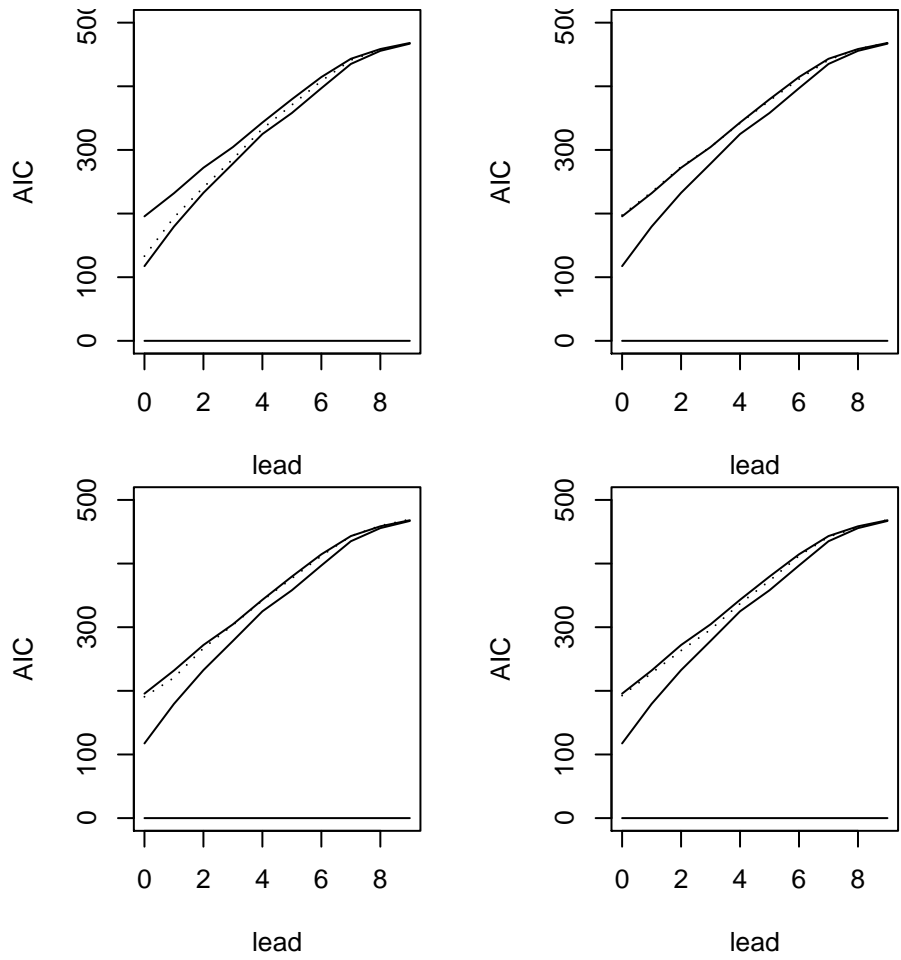}
  \end{center}
  \caption{
The two solid lines from figure~\ref{f:f1}, along with AIC scores for four more models (dotted lines):
a) seasonal bias, b) seasonal damping, c) seasonal noise and d) non-seasonal spread regression.
          }
  \label{f:f2}
\end{figure}

\begin{figure}[!htb]
  \begin{center}
    \includegraphics{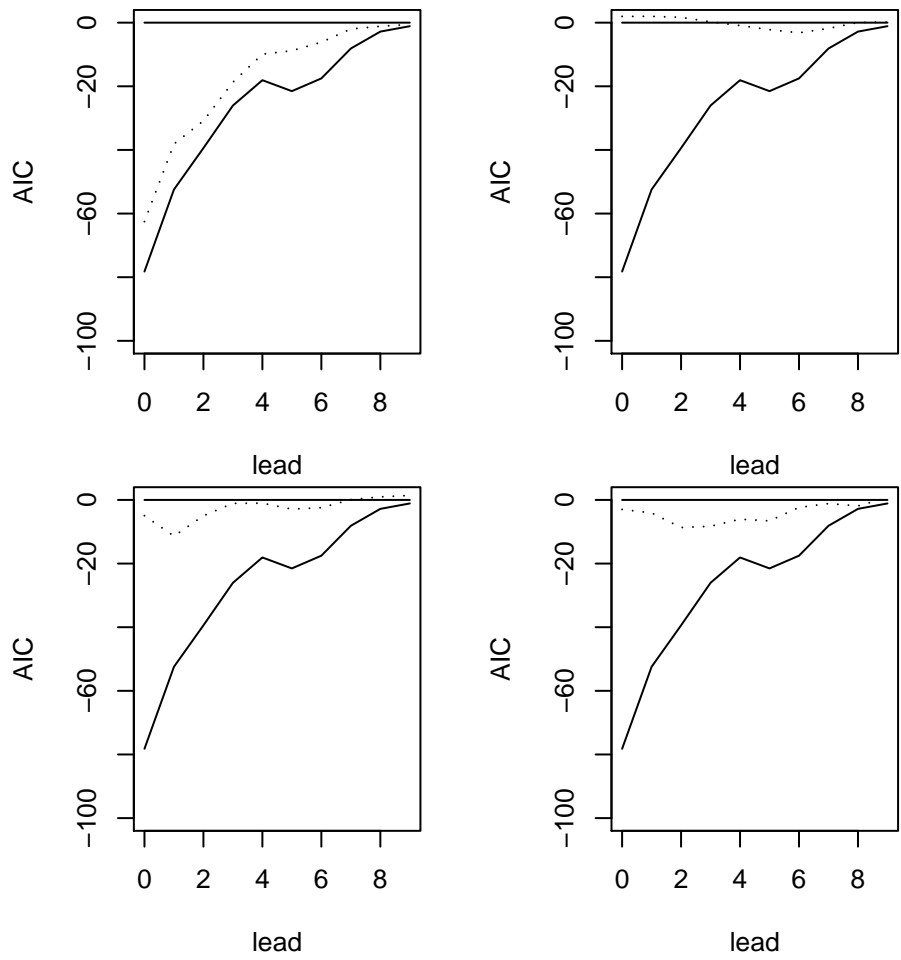}
  \end{center}
  \caption{
As figure~\ref{f:f2}, but all values shown relative to the AIC score for linear regression.
          }
  \label{f:f3}
\end{figure}

Where does this vast improvement come from? Is it the seasonality in the 
bias correction, the damping, the innovations, or all three in combination together? 
We address this question using the intermediate models.
Figure~\ref{f:f2} shows the same results as figure~\ref{f:f1} in the solid lines in each panel,
but also shows the AIC score for 4 other models as a dotted lines (the top left panel is the seasonal
bias model, top right is the seasonal damping model, lower left is the seasonal noise model and lower right is the
non-seasonal spread regression model). We immediately see that, of these models, 
the seasonal bias model gives the greatest
benefit over straight linear regression. Figure~\ref{f:f3} shows the same data, but now relative
to the AIC score of linear regression (more negative values are better). 
We see again that the seasonal bias correction gives a large
benefit, the seasonal damping gives no real benefit, the seasonal noise gives a small benefit
at short leads, and that the spread regression gives a small benefit at all leads. 

\begin{figure}[!htb]
  \begin{center}
    \includegraphics{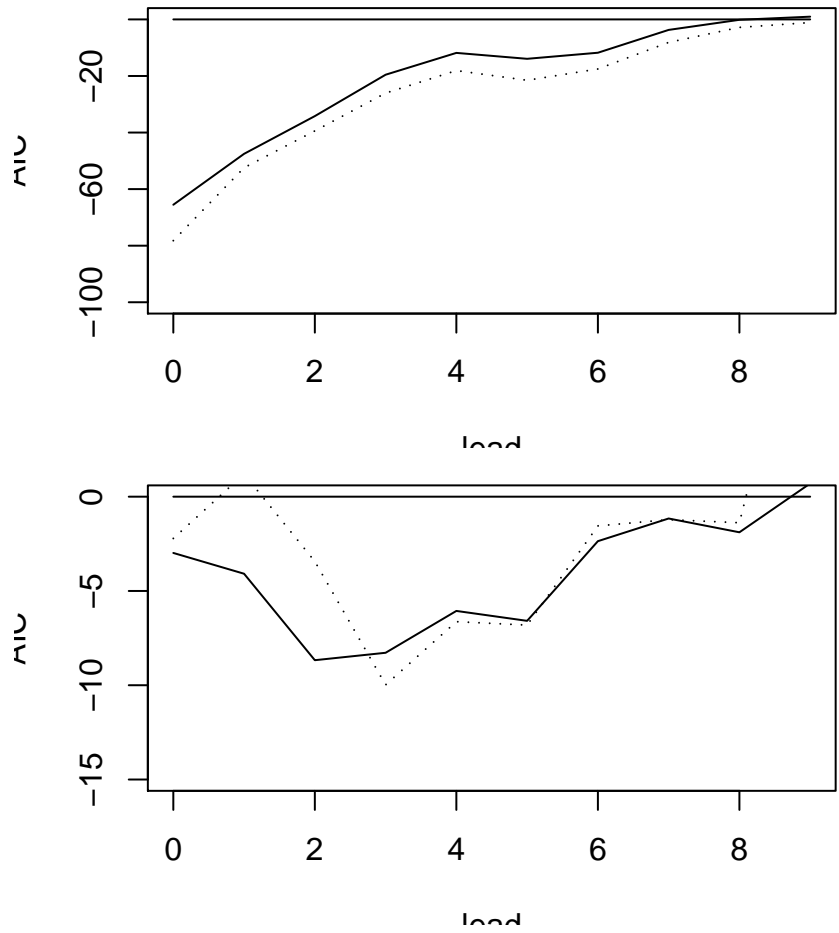}
  \end{center}
  \caption{
The synergy among the seasonal regression parameters, and between seasonality and spread regression.
          }
  \label{f:f10}
\end{figure}

Figure~\ref{f:f10} investigates the extent to which there is \emph{synergy} between the different
parameterisations. In the upper panel we compare the benefit from making all of $\alpha, \beta$ and $\gamma$
seasonal at once (dotted line) against the sum of the benefits of making them seasonal separately (solid line). 
There is definitely
some synergy: the total is greater than the sum of the parts. In the lower panel we consider the benefit
of using spread regression in a non-seasonal and seasonal model. In our previous research it
has been rather disappointing that using the ensemble spread brings so little benefit to probabilistic
forecasting, and we had hoped that maybe the benefit would be greater in the context of seasonal 
parameters for the mean. However, this is not the case. The benefit is more or less the same and is still very small.

\begin{figure}[!htb]
  \begin{center}
    \includegraphics{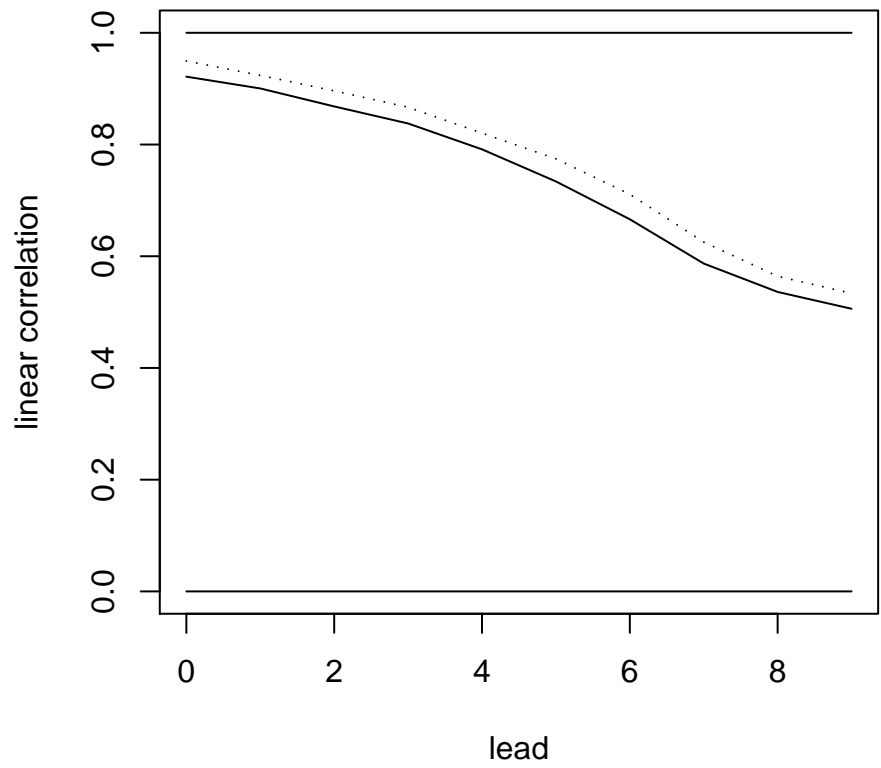}
  \end{center}
  \caption{
The linear correlation before (solid line) and after (dotted line) calibration with the seasonal regression model.
          }
  \label{f:f9}
\end{figure}

Constant parameter linear transformations such as the basic linear regression model cannot improve the
linear correlation of the ensemble mean with the observations. However, seasonal parameter linear
transformations can. Figure~\ref{f:f9} shows linear correlations before (solid line) and after (dotted line) the 
seasonal transformation. We see a definite improvement in linear correlation at all lead times.

\begin{figure}[!htb]
  \begin{center}
    \includegraphics{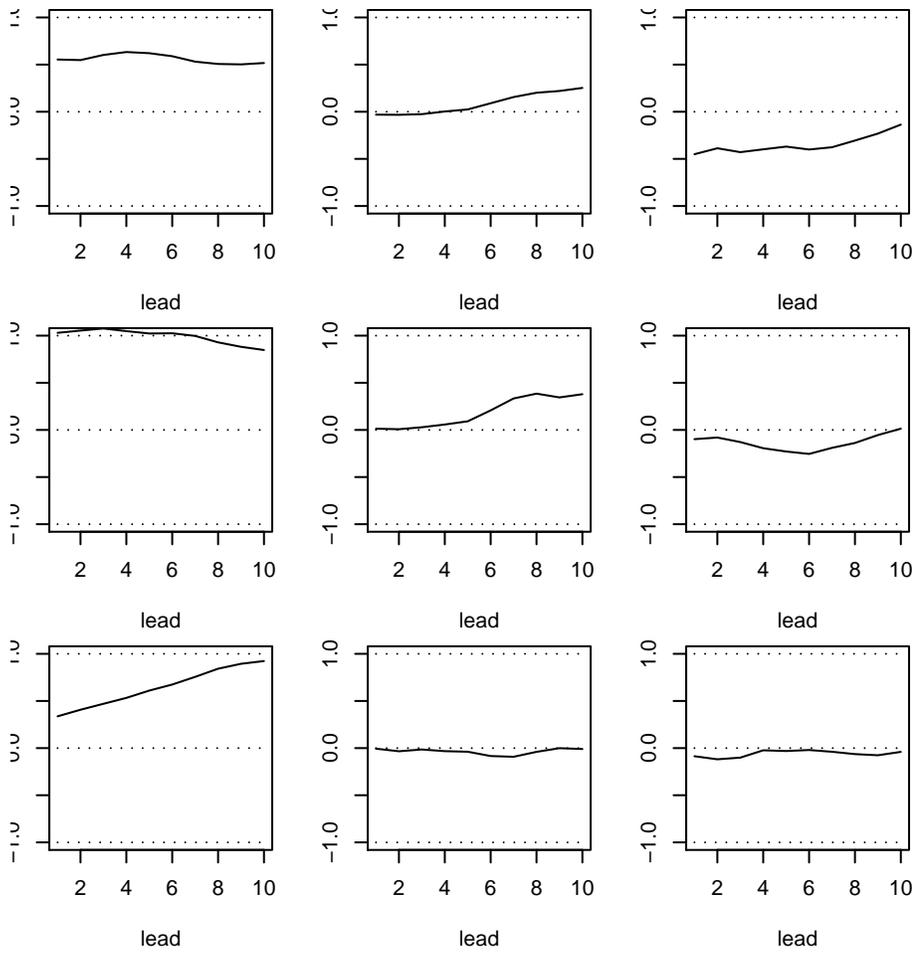}
  \end{center}
  \caption{
The 9 parameters of the seasonal regression model versus lead time.
The top row shows the alphas, the second row the betas and the third row
the gammas.
           }
  \label{f:f4}
\end{figure}

Figure~\ref{f:f4} shows the 9 parameters for the seasonal regression model. The top row shows the
alpha parameters, the second row the beta parameters and the third row the gamma parameters. 
We see that the variability in alpha is dominated by the cosine term at all but the longest leads,
the variability in beta is small relative to the average level of beta, and that gamma is more or
less constant throughout the year (but not with lead, of course).

%We have attempted to put confidence intervals on these parameter values, using the standard method of
%inverting the Fisher information matrix. However, this only worked for the three of the models:
%the linear regression model, the seasonal bias model and the spread regression model. For the other models
%(which include seasonal damping and noise parameters) the Fisher information matrix was not invertible.
%We suspect that this implies that these parameters are very poorly estimated. The uncertainty on the
%alpha and gamma parameters in the seasonal bias model is shown in figure~\ref{f:f5}, as two dotted
%lines either side of the parameter estimate. The parameters are so well estimated that these
%lines are almost indistinguishable for the parameter estimate itself.

\begin{figure}[!htb]
  \begin{center}
    \includegraphics{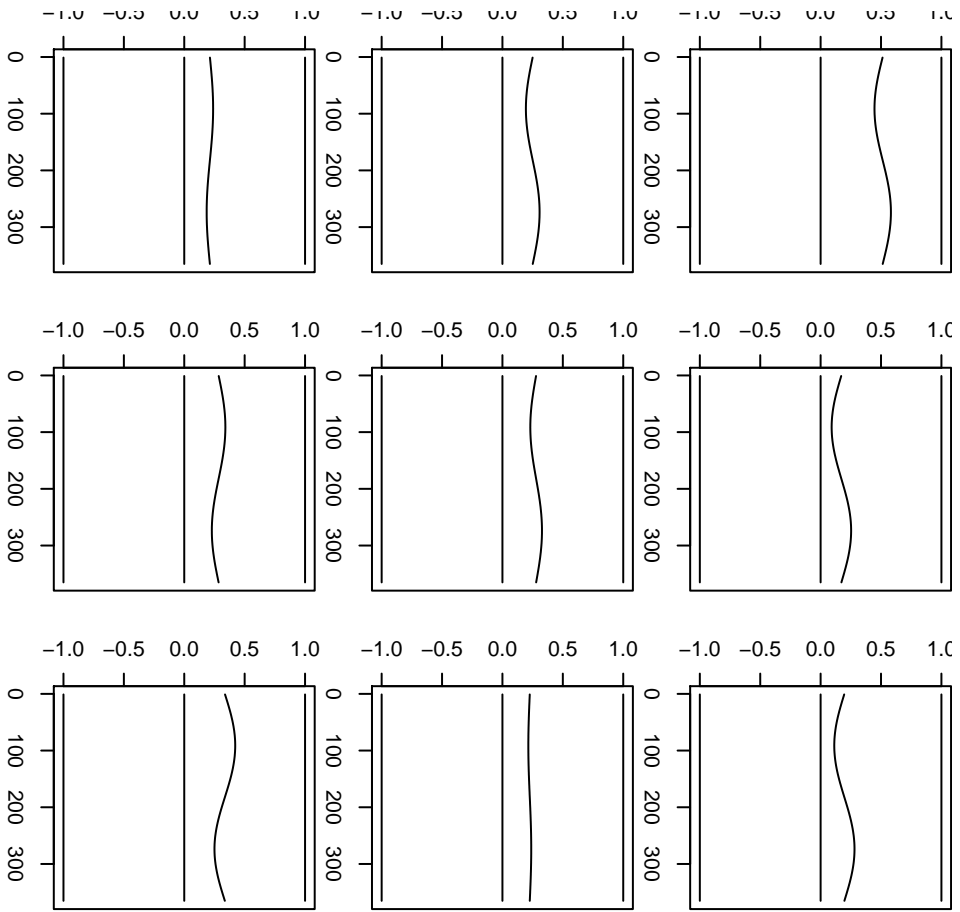}
  \end{center}
  \caption{
The seasonal variation of alpha predicted by the seasonal bias model, for leads
0 to 8.
          }
  \label{f:f6}
\end{figure}

The final figure, figure~\ref{f:f6}, shows the seasonal variability of alpha predicted by the seasonal bias model
for the first 9 lead times. We see that there is significant seasonal variability in the alpha predicted by the
model, which is consistent with the large effect that this model had on the AIC score. We see that
at short leads the smallest values of $\alpha$ are in spring and the largest in autumn, while at longer leads the 
opposite is true. 

\section{Summary}
\label{conc}

We have made \emph{another} attempt at improving the probabilistic forecasts of temperature that can be
made from ensemble forecasts. As before our starting point and basis for comparison is a linear regression model. 
Our previous attempts, that have looked at the benefit from using the ensemble spread 
(\citet{jewsonbz03a} and~\citet{jewson03g})
and the benefit from using the distribution of
the individual ensemble members~\citep{jewson03h}, have not shown much improvement over linear regression.
This time we have tried allowing the parameters of the regression model to vary seasonally. We find a \emph{dramatic}
improvement in the skill of the forecasts, much larger than the improvement from our previous attempts.
When we break down which terms are driving the improvement in skill we find that adding seasonality in the
bias is the most important factor. However seasonality in all three terms is important and there is actually
synergy between the terms such that the benefit from making all three regression parameters seasonal is greater than the
sum of the benefits of making each one seasonal separately (by the measure we use for skill).
Furthermore, our seasonal regression model also improves the linear correlation between forecast and observations.

The clear implication of this is that one should always 
use the seasonal parameter regression model in preference to the
constant parameter linear regression model, 
and that the seasonal parameter regression model should become the new baseline for comparison with other
methods and models.

There are, as ever, a number of avenues for future work that are suggested by this study.
Most obviously, allowing the bias to vary seasonally seems to be so important that 
one could try using more harmonics. It may well be that adding extra parameters in the modelling
of the bias is more useful than adding extra parameters elsewhere. We do not, however,
feel it would be justified with only the limited amount of data used in this study, and this is why we have not
considered higher harmonics here.

At a technical level, our fitting algorithm could be improved if we avoided the assumption that the 
forecast errors are uncorrelated in time (they are, in fact, weakly positively correlated). 
This may affect the results somewhat, but we doubt it would
affect them qualitatively. 

One possible criticism of our study might be that, by using up to 12 parameters to model only 
365 (weakly correlated) observation pairs we are flirting with overfitting. We wouldn't disagree.
We have compensated for this by using AIC rather than straight log-likelihood as our measure of skill 
and so it should be the case that our results would transfer to out of sample likelihood comparisons.
Nevertheless if longer data sets of stationary past forecasts ever become available
then it would be very interesting to repeat this analysis:
the parameters of the models we have presented will become much better estimated, and the results
that much better justified.

Our highest priority is now to repeat this analysis on wind and precipitation forecasts, which 
present similar but different challenges to the modelling of temperature because they are not close to 
normally distributed.

\section{Acknowledgements}

Many thanks to Ken Mylne and Caroline Woolcock for providing the forecast data used in this study, and for
helpful discussions.

\section{Legal statement}

The lead author was employed by RMS at the time that this article was written.

However, neither the research behind this article nor the writing of this
article were in the course of his employment,
(where 'in the course of his employment' is within the meaning of the Copyright, Designs and Patents Act 1988, Section 11),
nor were they in the course of his normal duties, or in the course of
duties falling outside his normal duties but specifically assigned to him
(where 'in the course of his normal duties' and 'in the course of duties
falling outside his normal duties' are within the meanings of the Patents Act 1977, Section 39).
Furthermore the article does not contain any proprietary information or
trade secrets of RMS.
As a result, the lead author is the owner of all the intellectual
property rights (including, but not limited to, copyright, moral rights,
design rights and rights to inventions) associated with and arising from
this article. The lead author reserves all these rights.
No-one may reproduce, store or transmit, in any form or by any
means, any part of this article without the author's prior written permission.
The moral rights of the lead author have been asserted.

\bibliography{jewson}

\newpage

\newpage

\newpage

\newpage

\newpage

\newpage

%\newpage
%\begin{figure}[!htb]
%  \begin{center}
%    \includegraphics{figs/fig5.ps}
%  \end{center}
%  \caption{
%The first four parameters (the alphas and the gamma) for the seasonal bias model.
%          }
%  \label{f:f5}
%\end{figure}

\newpage

\end{document}